\documentclass[twocolumn,showpacs, preprintnumbers,amsmath,amssymb,aps]{revtex4}
\usepackage{bm}
\usepackage{latexsym}
\usepackage{eufrak}
\usepackage[latin1]{inputenc}
\usepackage{latexsym}
\usepackage{amsmath}
\usepackage{epsfig}
\usepackage{ifpdf}
\usepackage{graphicx}
\usepackage{dcolumn}
\usepackage{amsmath}
\usepackage{amsfonts}
\usepackage{amssymb}
\usepackage{color}

\begin{document}
\title{Dark Energy as a Geometric Behavior with Lower Dimensionality: A Toy Model.}

\author{Miguel A. Garc\'{\i}a-Aspeitia \footnote{Part of the Instituto Avanzado de Cosmolog\'ia (IAC) collaboration http://www.iac.edu.mx/}} 
\email{agarca@fis.cinvestav.mx}
\affiliation{Divisi\'on de Ingenier\'ia Industrial y de Sistemas. Universidad Polit\'ecnica del Valle de M\'exico, Av. Mexiquense s/n, Col. Villa Esmeralda, C.P. 54910 Tultitl\'an, Estado de M\'exico. M\'exico.}
\affiliation{Departamento de F\'isica, DCI, Campus Le\'on, Universidad de Guanajuato, C.P. 37150, Le\'on, Guanajuato, M\'exico}

\begin{abstract}
In this paper, we investigate the possibility that a geometric model of dark energy that assume \textquotedblleft lower dimensions\textquotedblright, could generates a repulsive pressure which cause the current accelerated expansion of the Universe. We show how this geometrical model resemble as quintessence, cosmological constant and phantom energy during the cosmological evolution of the Universe choosing the corresponding matter scale factor as well as the possible consequences of this model in the Universe fate. Finally we perform the dynamical evolution of the model with the other Universe components as well as its domination epoch.
\end{abstract}

\draft
\date{\today}
\maketitle

\section{Introduction.}

Several cosmological observations at high redshift with supernovae of the Type Ia \cite{Perlmutter}, \cite{Riess} show evidence of the current accelerated expansion of the Universe. Similarly, observations of anisotropies of cosmic microwave background radiation (CMBR) \cite{Pope} suggest that the responsible of this acceleration is an entity known as dark energy whose main characteristic is that the equation of state (EoS) satisfies $\omega<-1/3$ and is not fulfilled by baryons, radiation (neutrinos and photons) neither dark matter and makes up $\sim70\%$ of the total content of the Universe \cite{Ferramacho}.

Currently, different proposals stress as candidates of dark energy the quintessence energy \cite{zlatev}, cosmological constant \cite{Carroll} \& \cite{Weinberg}, phantom energy \cite{caldwell} and other several candidates that comes from higher dimensional theories \cite{Miguel1} \& \cite{Miguel2} which satisfy the suitable EoS for an accelerated expansion or that provides the ability of accelerate the Universe by means of topological effects (for a review, see e.g. \cite{Wang}). It is important to remark that among all these proposals, the cosmological constant is the most accepted candidate for dark energy.

However, the best candidate is \emph{plagued} of fundamental problems among which are the fine tuning between the observational results which predict an energy density of $\rho_{_{\Lambda}}^{obs}<(10^{-12}GeV)^{4}$ in contrast with the theoretical results that predict $\rho_{_{\Lambda}}^{theo}\sim(10^{18}GeV)^{4}$ \cite{Carroll}, \cite{Weinberg} (120 orders of magnitude) as well as the coincidence problem associated with the recent Universe acceleration.

Focusing on solve the problems of dark energy nature; this paper proposes a geometric lower dimensional solution postulating that \emph{fundamental regions} permeate the Universe in a homogeneous and isotropic distribution. We consider that each fundamental region is in its minimum energy state related with the Heisenberg uncertainty principle as well as the relation of the size of this regions with the Schwarzschild radius and the consideration of a relativistic behavior as an extension of the results obtained by several authors \cite{Lee}-\cite{Li}. Remarkably, the counting of this fundamental regions must be performed only in one preferential direction and not in the three space dimensional directions related with the space-time structure; suggesting that at high energies, the Universe could behave as $(1+1)$ dimensional manifold as is argued by several authors (see e.g. \cite{Anch}-\cite{Modesto}). It should be noted that these kind of cosmological models with $(1+1)$ space-time dimensions, provide an interesting theoretical \textquotedblleft laboratory\textquotedblright to explore the behavior of classical and quantum gravity \cite{Chan}. The same calculation performed in the three dimensions immediately generates a wrong value of the energy density which can be interpreted as the cosmological constant fine tuning problem. 

Interestingly the previous conditions generates a natural accelerated expansion of the Universe when the scale factor corresponds to the matter domination era. Furthermore, the resulting EoS evolves like quintessence, cosmological constant and finally as phantom energy; having the remarkable goal that the energy density behaves as that predicted by the holographic principle \cite{Fischler} where the information is enclosed in the area of the physical object.
Finally, using a constriction to reduce the calculations, we analyze the dynamical system obtained from the Friedmann and Raychaudhuri equations for this Universe with the other components (dark matter, baryons, photons and neutrinos).
 We will henceforth use units in which $c=\hbar=1$.
 
\section{Fundamental Geometric Regions.}

\noindent \emph{Postulates}. We considered a Fundamental Geometric Regions (FGR) with relativistic energy $E=\gamma m_{0}$ and linear momentum $\textbf{p}=m_{0}\textbf{v}\gamma$, where $\gamma=(1-\textbf{v}^{2})^{-1/2}$ is the Lorentz factor, $m_{0}$ is the rest mass of the fundamental region and $v$ is the velocity associated with the FGR. It is important to remark that the relativistic treatment generates new dynamics in high energy environments in comparison with a no relativistic treatment. This new scope is important for the understanding of the FGR and its relation with the elusive dark energy.  

In the same way the position and momentum of the FGR in its minimal state is determined through the Heisenberg principle as $\Delta x\Delta p_{x}\simeq1/2$.

Then, the combination of these postulates lead to the following expression $xv\gamma m_{0}\simeq1/2$, where we have assumed the following association: $\Delta x\to x$ and $\Delta p_{x}\to p_{x}=m_{0}v\gamma$ \cite{Lattman}. 

If the length of the FGR is related with the Schwarzschild diameter as $x=4Gm_{0}$ \cite{Lattman} where $G$ is the Newtonian constant, then the energy and length of these structures can be written as

\begin{equation}
\epsilon\simeq\sqrt{\left(\frac{m_{p}^{2}}{8v}\right)\gamma}, \;\; x\simeq\sqrt{\left(\frac{2}{vm_{p}^{2}}\right)\gamma^{-1}}, \label{2}
\end{equation}
where $m_{p}$ is the Planck mass and it is related with the Newtonian constant as $m_{p}^{2}=G^{-1}$. If the FGR is immersed in a homogeneous and isotropic distribution of the Universe, we can accept that the total energy read as

\begin{equation}
\varepsilon(t)_{_{T}}^{^{(1+1)D}}\simeq \sum_{i}^{n}\epsilon=\left\lbrace\frac{2X(t)}{x}\right\rbrace\epsilon, \label{energytot}
\end{equation}
where $n$ is the number of FGR in a $(1+1)D$ region of space in concordance with the expectations of the \textquotedblleft lower dimensions\textquotedblright \; models \cite{Anch}, \cite{Mureika} (it is possible to demonstrate that if the three dimensions of the space $n^3$ is added, the value of the energy density is not in agreement with the observational result \footnote{See Appendix \ref{AppendixA} for a demonstration}), $X(t)$ the physical radius of the Universe (thus $2X(t)$ is the physical diameter of the Universe). Using \eqref{2} in \eqref{energytot} the total energy as a result of the contributions of each FGR in all the $(1+1)D$ Universe is $\varepsilon(t)_{_{T}}^{^{(1+1)D}}\simeq m_{p}^{2}\gamma X(t)/2$.

On the other hand, observations confirm that the Universe behave as a $(3+1)$ dimensional manifold; then it is necessary project the $(1+1)$ dimensional results in the $(3+1)$ dimensional Universe. Then, it is possible to expect that the volume can be written as $V(t)\sim4\pi X(t)^{3}/3$ \cite{Holo}, this is because flatness implies that the geometry of the hypersurface is Euclidean in $\mathbb{R}^{3}$. Therefore it is possible to define the volume of the Universe as the usual of a three sphere $\mathbb{S}^{3}$. Finally the $(1+1)D$ energy can be projected in the $(3+1)D$ Universe with the equation $\rho\equiv \varepsilon^{^{(1+1)D}}_{_{T}}/V$ and then the energy density of the FGR can be written as

\begin{equation}
\rho(t)_{_{Y}}\simeq\frac{3}{4\pi X(t)^{3}}\sum_{i=1}^{2X(t)/x}\epsilon=\frac{3m_{p}^{2}}{8\pi}\frac{(1-\dot{X}(t)^2)^{-1/2}}{X(t)^{2}}, \label{eq3}
\end{equation}
which coincide with the  holographic principle, due that the $Area\sim X^{2}$ \cite{Bousso}-\cite{Zhang}. It should be noted that $\dot{X}(t)\sim v$ under the assumption that the velocity of the FGR its part of the space-time.  The idea behind the cut-off scale can be described with the following words: A minimal information should exist and the consistent density associated to this minimal counterpart should be employed as an energy density.

With the aim of study the dynamics in a cosmological context, a useful expression of the last equation can be written in a comoving coordinates $\textbf{X}(t)=a(t)\textbf{x}^{\prime}$, in the following way

\begin{equation}
\rho(t)_{_{Y}}\simeq\rho_{0}a(t)^{-2}\left[1-x^{\prime2}\dot{a}(t)^{2}\right]^{-1/2}, \label{rho}
\end{equation}
where the velocity of this regions is associated with the rate of the expansion of the Universe as $\dot{\textbf{X}}(t)\simeq\dot{a}(t)\textbf{x}^{\prime}$. Thus $\rho_{0}$ can be defined as $\rho_{0}=3m_{p}^{2}/8\pi x^{\prime2}$, $a(t)$ is the scale factor of the Universe and $\textbf{x}^{\prime}$ is the distance in a comoving reference system which is the free parameter associated to the theory. 

Considering a fluid behavior we accept the following equation $\dot{\rho}_{_{Y}}+(3\dot{a}(t)/a(t))(P_{_{Y}}+\rho_{_{Y}})=0$. Then the pressure of the FGR, can be written as

\begin{eqnarray}
P(t)_{_{Y}}&&\simeq-\frac{\rho_{0}}{3}\left[a(t)^{-2}+x^{\prime2}\frac{\ddot{a}(t)}{a(t)}(1-x^{\prime2}\dot{a}(t)^{2})^{-1}\right]\times\nonumber\\&&(1-x^{\prime2}\dot{a}(t)^{2})^{-1/2}. \label{p}
\end{eqnarray}
It should be noted that fixing $x^{\prime}$, the equations \eqref{rho} and \eqref{p} are subject to the constriction $x^{\prime}\dot{a}(t)<1$. Finally, the EoS can be expressed as

\begin{eqnarray}
\omega(t)_{_{Y}}&&\simeq-\frac{1}{3}\left[\frac{1+x^{\prime2}a(t)^{2}\dot{H}(t)}{1-x^{\prime2}a(t)^{2}H(t)^{2}}\right], \label{omega}
\end{eqnarray}
where $\omega(t)_{_{Y}}$ can be written in terms of the Hubble parameter $H(t)\equiv\dot{a}(t)/a(t)$ and the scale factor. It is important to note that in the equation \eqref{omega} we use the relation $P_{_{Y}}\equiv\omega_{_{Y}}\rho_{_{Y}}$.

\section{Testing $\omega_{_{Y}}$ in the Case of a Matter Domination Universe.} 

We analyze the behavior of the EoS when the scale factor corresponds to the matter domination era in order to extract information about its evolution. Thus, the scale factor evolves as $a(t)=\vartheta t^{2/3}$ where $\vartheta=(6\pi m_{p}^{-2}\rho_{m})^{1/3}$ and $\rho_{m}$ is the matter density. Then, the equation \eqref{omega} can be written as

\begin{equation}
\omega(z)_{_{Y}}\simeq-\frac{1}{3}+\frac{1}{\frac{27}{2}\eta^{2}\left(\frac{1}{z+1}\right)-6}, \label{evalmatt}
\end{equation}
where $\eta=1/x^{\prime}\vartheta^{3/2}$ and $z$ is the redshift that it is related to the scale factor as $1+z=a^{-1}$. 

Following Fischler \emph{et al.} \cite{Fischler}, we assume $x^{\prime}\equiv R_{H}(t_{0})$ as the particle horizon used in the holographic cosmology (Li propose the future event horizon \cite{Li} in a no relativistic limit), thus $x^{\prime}$ can be expressed with the following equation

\begin{equation}
x(t_{0})^{\prime}\equiv R_{H}(t_{0})=a(t_{0})\int_{0}^{t_{0}}\frac{dt^{\prime}}{a(t^{\prime})}.
\end{equation} 
For matter domination we obtain that $x(t_{0})^{\prime}=3t_{0}$ where $t_{0}$ is the current age of the universe ($t_{0}\sim4.3\times10^{17}s$), then $x^{\prime}\sim3.87\times10^{26}m$ (where it is temporarily re-introduced the units).

In term of time $t^{2/3}=1/\vartheta(1+z)$, the previous expression \eqref{evalmatt} tell us the following aspects:

\renewcommand{\labelenumi}{\Roman{enumi}.}

\begin{figure}[htp]
\centering
\includegraphics[scale=0.38]{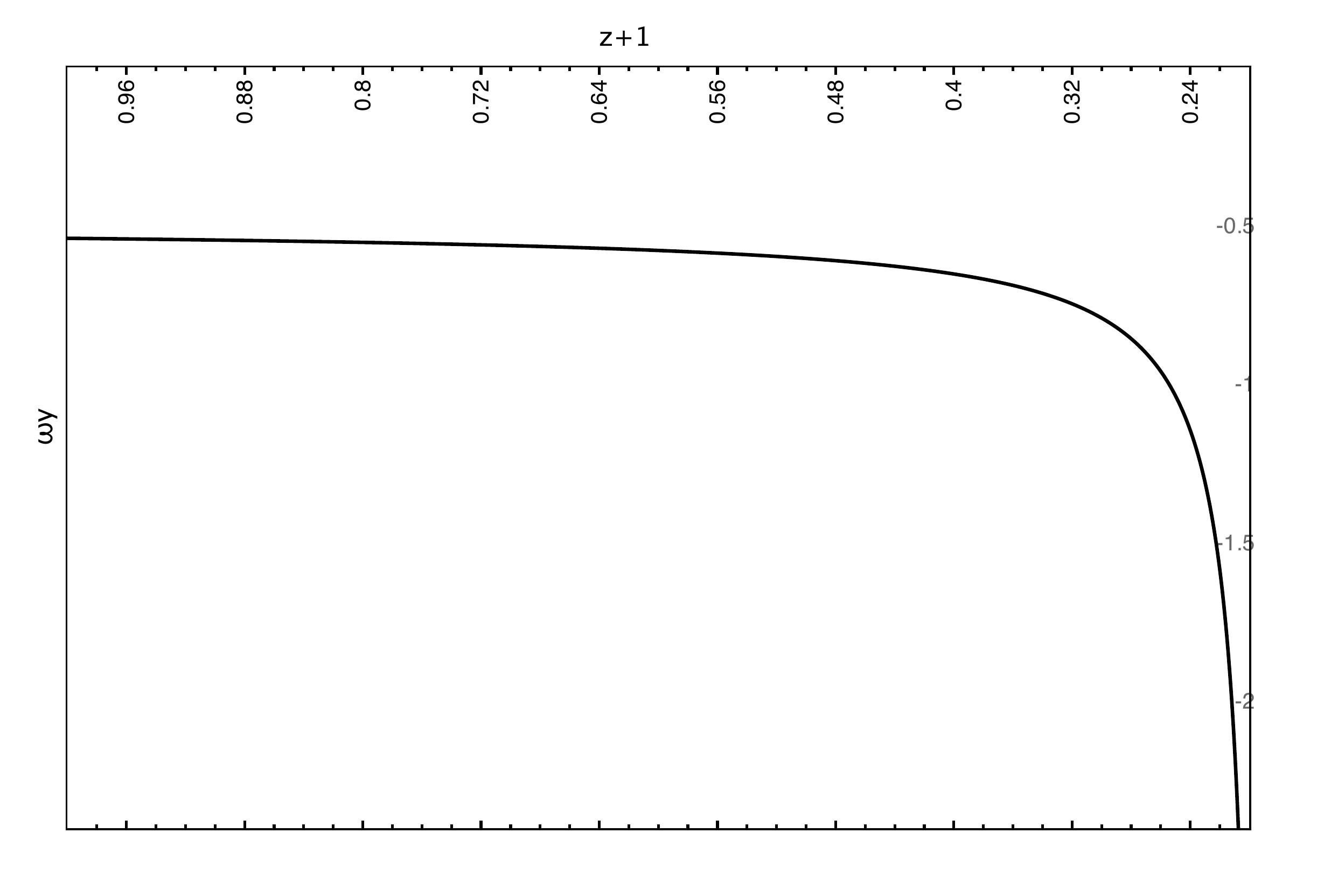}
\caption{Evolution of the EoS $\omega_{_{Y}}$ \emph{vs} redshift $z+1$ from the equation \eqref{evalmatt}. The EoS begins with $\omega_{_{Y}}\sim-0.5$ and diverge when the redshift grows $z+1$.}
\label{figure}
\end{figure}

\begin{enumerate}

\item Under those hypothesis; currently, the EoS is $\omega(t_{0})_{_{Y}}\sim-0.6$ which corresponds to quintessence models ($-1<\omega<-1/3$) \cite{zlatev} (See figure \ref{figure}).

\item The EoS for cosmological constant $\omega_{Y}=-1$ is at $t\sim1.515\times10^{18}s$ (See figure \ref{figure}) \cite{Carroll}-\cite{Weinberg}.

\item And after this era, the phantom energy ($\omega<-1$) \cite{caldwell} dominates the scene triggering a Big Rip at $\sim60Gyrs$ after this moment; this is because $\omega_{_{Y}}$ tends to diverge as $t$ grows (See figure \ref{figure}). For example Caldwell \emph{et al.} argues that the remaining time for the Big Rip with phantom energy as source of dark energy, is at $\sim35Gyrs$ (see e.g. \cite{caldwell}).

\end{enumerate}

This peculiar behavior suggests that the FGR could be considered as \emph{source of dark energy}.

It is important to stress that the constriction generate that the differentiation of the scale factor must be $\dot{a}(t)<7.751\times10^{-19}s^{-1}$. 

\section{Dynamical Analysis with a Constriction.}

We study the dynamical evolution of the Universe which can be described with the Einstein field equations $G_{\mu\nu}=8\pi m_{p}^{-2}T_{\mu\nu}$ where the energy momentum tensor can be written as 

\begin{equation}
T_{\mu\nu}=\sum_{i,other}\left(T_{\mu\nu}\right)_{i}+diag(-\rho_{_{Y}},\vec{P}_{_{Y}}),\label{einst}
\end{equation}
being the first term in the equation \eqref{einst} the corresponding to the matter (dark matter and baryons), radiation (photons and neutrinos) and the \eqref{einst} corresponds to FGR (we have used the signature $diag(-+++)$).

Applying the Einstein field equations, we obtain the following Friedmann and Raychaudhuri equations as

\begin{equation}
H^{2}=\frac{\kappa^{2}}{3}\left(\rho_{m}+\rho_{_{\Gamma}}+\rho_{_{Y}}\right), \label{Friedm}
\end{equation}

\begin{equation}
\dot{H}=-\frac{\kappa^{2}}{2}\left[\rho_{m}+\frac{4}{3}\rho_{_{\Gamma}}+\left\lbrace1-\frac{1+x^{\prime2}a^{2}\dot{H}}{3(1-x^{\prime2}a^{2}{H}^{2})}\right\rbrace\rho_{_{Y}}\right], \label{raych}
\end{equation}
being $\kappa^{2}\equiv8\pi G\equiv8\pi m_{p}^{-2}$, $(m)$ is the indication of matter composed by dark matter and baryons $(dm,b)$, ($\Gamma$) is the indications of radiation composed by photons and neutrinos $(\gamma,\nu)$; finally ($Y$) is for the \emph{dark energy} of FGR proposed in the model, where $H\equiv H(t)$ and $a\equiv a(t)$. Equivalently the conservation equation for each Universe component can be written as

 \begin{subequations}
\begin{eqnarray}
{\dot\rho_{m}}&+&3\,H \rho_{m}=0,\\
{\dot\rho_{_{\Gamma}}}&+& 4\,H \rho_{_{\Gamma}}=0, \\
{\dot\rho_{_{Y}}}&+& \,3H\rho_{_{Y}}=H\left(\frac{1+x^{\prime2}a^{2}\dot{H}}{1-x^{\prime2}a^{2}{H}^{2}}\right)\rho_{_{Y}},
\end{eqnarray}\label{eq:back}
\end{subequations}
assuming that $\partial T^{\mu}_{\nu}/\partial x^{\mu}=0$. The equation (\ref{eq:back}e) is significantly different to the other conservation equations due to the source term in the right hand of the expression. This source term generates the accelerated expansion expected in the dynamic of the Universe. Similarly, this behavior can be observed in the Raychaudhuri equation \eqref{raych} where $\dot{H}$ becomes positive.

In order to solve the system of equations \eqref{Friedm}-\eqref{raych} we define the following dimensionless variables

\begin{eqnarray}
y^2\equiv\frac{\kappa^2}{3}\frac{\rho_{dm}}{H^2}, \quad&
z^2&\equiv\frac{\kappa^2}{3}\frac{\rho_{b}}{H^2},\nonumber\\
l^2\equiv\frac{\kappa^2}{3}\frac{\rho_{\gamma}}{H^2},\quad&
m^2&\equiv\frac{\kappa^2}{3}\frac{\rho_{\nu}}{H^2},\nonumber\\
x^2\equiv\frac{\kappa^2}{3}\frac{\rho_{_{Y}}}{H^2},
\label{eq:varb}
\end{eqnarray}
where it is remarked explicitly all the Universe components and it is assumed the following constriction $H^{2}=(\kappa^{2}/3)\rho_{_{Y}}$ in \emph{order to simplify the equations for FGR}. 

Using these variables, the equations for the evolution of the background Universe can be transformed into the following equations

\begin{subequations}
\begin{eqnarray}
y'&=& \frac{3}{2}\left(\Pi-1\right)y, \label{eq:dsb_u}\\
z'&=& \frac{3}{2}\left(\Pi-1\right)z, \label{eq:dsb_v}\\
l'&=&\frac{3}{2}\left(\Pi -\frac{4}{3}\right)l, \label{eq:dsb_b}\\
m'&=&\frac{3}{2}\left(\Pi-\frac{4}{3} \right)m, \label{eq:dsb_z}\\
x'&=&\frac{3}{2}\left(\Pi-\frac{2}{3}\right)x, \label{eq:dsb_x}
\end{eqnarray}\label{eq:dsb}
\end{subequations}
where the prime denotes a derivative with respect to the e-folding number $N=ln(a)$, and $\Pi$ is defined as

\begin{equation}
\frac{3}{2}\Pi\equiv-\frac{\dot{H}}{H^{2}}=\frac{3}{2}\left[y^2+z^2+\frac{4}{3}l^2+\frac{4}{3}m^2+\frac{2}{3}x^2\right], \label{17}
\end{equation}
subject to the Friedmann constraint

\begin{equation}
y^2+z^2+b^2+m^2+x^2=1.
\end{equation}
To solve the dynamical system \eqref{eq:dsb_u}-\eqref{17} it is necessary impose the initial conditions to the $\Omega_{i}$ taken from $7$-years WMAP values (See Table \ref{TableI}) and the model prediction for dark energy as $\Omega_{_{Y}}\equiv\rho_{Y}/\rho_{crit}\simeq0.73$.

\begin{table}[!hbt]
\begin{center}
\begin{tabular}{|l|l|}
\hline
Density\; Parameter &  $WMAP \; 7$ $ML^{*}$\\
\hline
$\Omega_{dm}^{0}$ & $0.226$\\
$\Omega_{b}^{0}$ & $0.0451$\\
\hline
\end{tabular} \caption{WMAP 7-year values. \textquotedblleft ML\textquotedblright refers to the Maximum Likelihood parameters \cite{WMAP7}} \label{TableI} \end{center} 
\end{table}
The density parameter for photons is $\Omega_{\gamma}^{0}\equiv\rho_{\gamma}/\rho_{crit}\simeq0.00004$ where $\rho_{\gamma}\sim T_{_{CMBR}}^{4}$. Equivalently, the density parameter for neutrinos can be written as $\Omega_{\nu}^{0}\simeq0.00002$ \cite{WMAP7}-\cite{Tegmark}.

\begin{figure}[htp]
\centering
\includegraphics[scale=0.5]{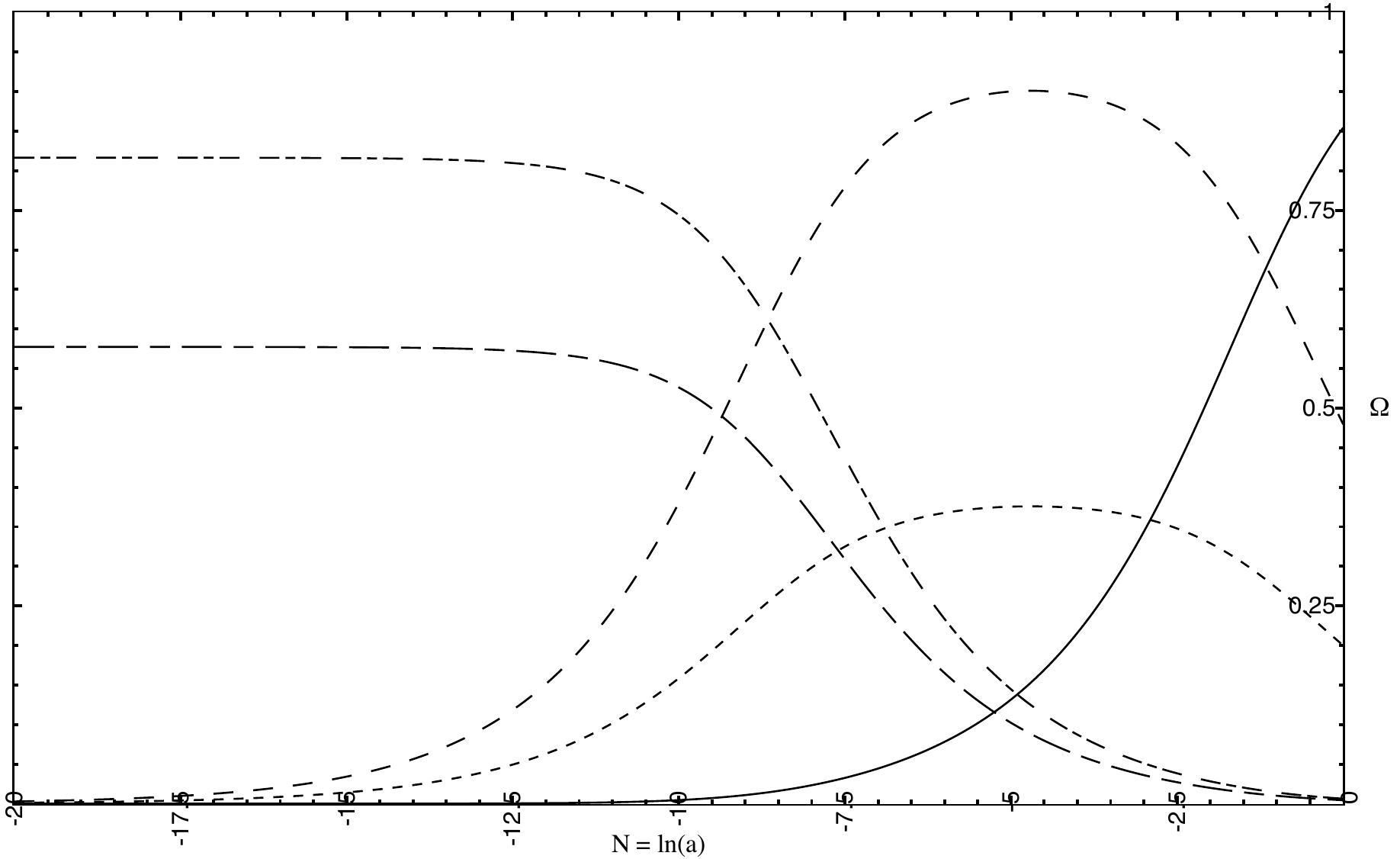}
\caption{Dynamic evolution of photons (dotted and dashed line), neutrinos (small and large dashed line), dark matter (large dashed line), baryons (small dashed line) and FGR which behave as dark energy (continuous line).}
\label{figure3}
\end{figure}
As we observe from the figure (\ref{figure3}) the constriction ($H^{2}=(\kappa^{2}/3)\rho_{_{Y}}$) generates that the evolution of the dark energy is subdominant in early times of the Universe allowing nucleosynthesis in radiation (photons$+$neutrinos ) domination era and the structure formation in the matter domination era. Correspondingly, the dark energy (composed with FGR) is dominant in later times in Universe evolution which fits with the observational results \cite{Perlmutter}-\cite{Pope}.

\section{Discussion.}

The most notable departure from this model is that the addition of relativistic corrections contributes to extra information of the behavior of this \emph{dark energy} model, introducing new dynamics which lack the Newtonian limit \cite{Lee}, \cite{Li}. To summarizing the results we remark the most important features of this model in the following way:

The EoS behave as quintessence, cosmological constant and phantom energy in different epochs of the Universe and predicts a Big Rip in $60Gyrs$ due to the divergence in the EoS (See figure (\ref{figure})) if it is choose a scale factor that corresponds to the matter scale factor in the equation \eqref{omega}. 

It should be noted that the addition of this relativistic correction give us a natural acceleration without the need of future horizon \cite{Lee}, \cite{Li}.

The supposition of the FGR and the reduction of the dimensions assumed in the \textquotedblleft lower dimensional models\textquotedblright, generates the correct value for the energy density of vacuum and the coincidence problem, giving us new clues about the nature of \emph{dark energy} and the accelerated expansion of the Universe. Doing emphasis that the results are in concordance with the holographic principle \cite{Fischler}. 

Additionally, we analyze the Friedmann, Raychaudhuri and conservation equations through a dynamical system showing that the FGR could be a strong candidate to be the source of dark energy due to that it is dominant in later times of the Universe evolution preserving nucleosynthesis and structure formation epoch. 

Better analysis of the EoS \eqref{omega} can be studied if we not use the constriction $H^{2}=(\kappa^{2}/3)\rho_{_{Y}}$ to simplify the dimensionless variables. In this study it is necessary ameliorate the dynamical equations together with the computational environment.

Further studies will explore this alternative to dark energy nature to obtain new clues of the behavior and fate of $\sim70\%$ of our Universe \cite{Luis} \& \cite{Durmos}, principally in the inflation epoch and in the extreme gravitational environments as galactic nuclei and stellar black holes. In a similar vein, it is necessary an exhaustive study of the hypothesis that the Universe behaves as $(1+1)$ dimensional space-time at Planck scales with the aim of found evidences of this particular behavior together with a most realistic scenario to understand the projection of the $(1+1)D$ to the $(3+1)D$ current Universe.

\section{Acknowledgment.}

I would like to acknowledge the help and the enlightening conversation with Yasm\'in Alc\'antara, Abdel P\'erez-Lorenzana, Juan Maga\~na, Eloy Ay\'on and Ivan Rodriguez. Instituto Avanzado de Cosmolog\'ia (IAC) collaboration.

\appendix

\section{The $n^3$ consideration for the $(3+1)D$ manifold.} \label{AppendixA}

Here we demonstrate that the consideration of $n^3$ due to the three dimensional space generate a wrong value known as \emph{fine tuning problem} of the density energy of vacuum. This is because the expression

\begin{equation}
\rho_{_{Y}}^{(3+1)D}\simeq\frac{3}{4\pi X(t)^3}\prod_{k=1}^{3}\left\lbrace\sum^{2X(t)/x}_{i=1}\right\rbrace_{k}\epsilon\simeq m_{p}^{4}, \label{A1}
\end{equation}
is equivalent to the integral equation proposed by Weinberg \cite{Carroll}, \cite{Weinberg}

\begin{equation}
\rho^{(3+1)D}=\int_{0}^{\Lambda}\frac{4\pi k^{2}dk}{(2\pi)^{3}}\frac{1}{2}\sqrt{k^{2}+m^{2}}\simeq\frac{\Lambda^{4}}{16\pi^{2}}\simeq m_{p}^{4}, \label{A2}
\end{equation}
where $\Lambda>>m$, and it is used the election of $\Lambda\simeq m_{p}/\sqrt{8\pi}$ then $\rho_{_{Y}}^{(3+1)D}\simeq\rho^{(3+1)D}\simeq m_{p}^{4}=2\times10^{71}GeV^{4}$. Conclusively, the choice of the equation \eqref{A1} generates the well known wrong \emph{constant} value of the density energy called \emph{the cosmological constant problem}.

\end{document}